\documentstyle[12pt,twoside]{article}
\def\greaterthansquiggle{\raise.3ex\hbox{$>$\kern-.75em\lower1ex\hbox{$\sim$}}}
\def\lessthansquiggle{\raise.3ex\hbox{$<$\kern-.75em\lower1ex\hbox{$\sim$}}}
\newcommand{\beq}{\begin{equation}}
\newcommand{\eeq}{\end{equation}}
\newcommand{\beqa}{\begin{eqnarray}}
\newcommand{\eeqa}{\end{eqnarray}}
\newcommand{\beqan}{\begin{eqnarray*}}
\newcommand{\eeqan}{\end{eqnarray*}}
\newcommand{\ba}{\begin{array}}
\newcommand{\ea}{\end{array}}

\newcommand{\A}{{\cal A}}

\newcommand{\M}{{\cal M}}

\newcommand{\st}{\stackrel}

\def\nz{\ifmmode {I\hskip -3pt N} \else {\hbox {$I\hskip -3pt N$}}\fi}
\def\zz{\ifmmode {Z\hskip -4.8pt Z} \else
       {\hbox {$Z\hskip -4.8pt Z$}}\fi}
\def\qz{\ifmmode {Q\hskip -5.0pt\vrule height6.0pt depth 0pt
       \hskip 6pt} \else {\hbox
       {$Q\hskip -5.0pt\vrule height6.0pt depth 0pt\hskip 6pt$}}\fi}
\def\rz{\ifmmode {I\hskip -3pt R} \else {\hbox {$I\hskip -3pt R$}}\fi}
\def\cz{\ifmmode {C\hskip -4.8pt\vrule height5.8pt\hskip 6.3pt} \else
       {\hbox {$C\hskip -4.8pt\vrule height5.8pt\hskip 6.3pt$}}\fi}
\newtheorem{theorem}{Theorem}

\def\au{{\setbox0=\hbox{\lower1.36775ex%
\hbox{''}\kern-.05em}\dp0=.36775ex\hskip0pt\box0}}
\def\ao{{}\kern-.10em\hbox{``}}

\normalsize
\begin{document}
\bibliographystyle{plain}

\begin{titlepage}
\begin{flushright}
\today
\end{flushright}
\vspace*{2.2cm}
\begin{center}
{\Large \bf Asymptotics for automorphisms of quantum systems on the lattice }\\[30pt]

Heide Narnhofer  $^\ast $\\ [10pt] {\small\it}
Fakult\"at f\"ur Physik \\ Universit\"at Wien\\

\vfill \vspace{0.4cm}

\begin{abstract}We study the behaviour of continuous automorphism groups of quantum spin systems on the lattice. Whereas the shift is norm asymptotically abelian continuous automorphism groups can lead only to delocalisation but not to norm asymptotic abelianess. As a consequence the shift does not allow a continuous extension.

\\
\hspace{1.9cm}

\end{abstract}
\end{center}

\vfill {\footnotesize}

$^\ast$ {E--mail address: heide.narnhofer@
univie.ac.at}
\end{titlepage}
\section{Introduction}
We are interested in the limit behaviour of continuous automorphism groups of quantum systems on the lattice. On this algebra the shift acts as an discrete automorphism group that is norm asymptotically abelian with finite speed. We want to see whether a similar behaviour is possible for continuous automorphism groups that commute with the shift and are approximately inner so that their derivative is given on the local algebras as a commutator with a local operator or an operator that is the norm limit of local operators. For these automorphisms it was shown that up to exponential approximation they share with the shift that they have finite speed \cite{LR}, \cite{Na}. We wonder whether they can also share the fact that they are norm asymptotically abelian and local operators escape to infinity.
We can compare with the Fermi algebra on the lattice where the shift is norm asymptotically abelian on the even algebra and norm asymptotically antiabelian on the odd algebra. In \cite{NI} we showed that continuous automorphism groups with some generality cannot be norm asymptotically abelian. An exception are quasifree automorphism groups. Especially interaction destroys asymptotic abelianess. On the other hand the evolution leads to delocalisation and slightly perturbed states of the tracial state (and probably of all invariant states) converge in the mean to the initial unperturbed state.
In \cite{AM} it was shown that most of the relevant properties of quantum systems on the lattice can be transferred to the Fermi system on the lattice, and vice versa. An exception is the shift, that for the Fermi system can be extended to a continuous automorphism group, but the applied constructions \cite{RWW} and \cite{MN} fail for the lattice system. Therefore it is worthwhile to examine how far the ideas and methods of \cite{NI} can be transferred to the lattice system.
As in \cite{NI} we consider the system in the tracial state. If abelianess fails here, it fails also in norm. In the GNS-construction asymptotic properties of the automorphism group can be read off from the asymptotic properties of the unitaries that implement the automorphisms. Further the commutant using the modular conjugation inherits the quasilocal structure of the observable algebra. Starting with the tensorproduct of  operators of the observable algebra located at a point combined with the corresponding operators of the commutant norm asymptotic abelianess implies that in the limit it commutes with the bounded operators in the Hilbertspace. For this tensorproduct we can construct with the help of the Weyl-operators in the matrix algebra a maximal abelian subalgebra. In this subalgebra an appropriate  onedimensional projector implements a maximally entangled state between the two matrix algebras. Under the action of the shift it converges strongly to $1$ for any automorphism group that is norm asymptotically abelian. Applying this idea to continuous automorphism groups that are approximated by inner automorphisms the corresponding derivative of the projection has to converge strongly to $0$. We will show that this leads to a contradiction.
As a consequence based on an argument in \cite{MN} the shift does not allow a continuous extension in the above sense, generalizing the results in \cite{RWW} and \cite{MN}.
Considering the question whether automorphisms lead to delocalisation the arguments in \cite{NI} do not refer to the Fermi structure, and we will repeat them with slight generalisations according to the matrices that can have dimensions different from 2.

\section{The extended algebra on the lattice in the tracial state}
We consider the observable algebra $\A =\otimes _{x\in Z^{\nu}}\M_x^d$ with $M_x^d$ the matrix algebra of dimension $d$ at the point $x$ as the quantum algebra over a lattice of dimension $\nu .$ With fixing an orthonormal basis $|s\rangle \quad s=1,,,d  $ in the corresponding Hilbertspace $H^d$ the matrix-algebra is spent by Weyl operators given by
$$ \langle s|U_{r_1,r_2}|s'\rangle =e^{\frac{2i\pi}{d}}\delta _{s-s'+r_2,0}.$$
They satisfy the algebraic relations $$U_{r_1,r_2}U_{t_1,t_2}=e^{\frac{2i\pi }{d}t_1r_2}U_{r_1+t_1,r_2+t_2}$$
and the corresponding commutation relations.

We represent the algebra in the GNS-representation given by the trace $\omega (AB)=\omega (BA)$ so that
$$\langle \Omega |\Pi (A) |\Omega \rangle =\omega (A).$$
This representation $\Pi (A)$ is cyclic and faithfull, i.e. every vector in the Hilbertspace can be approximated by $ \Pi (A)|\Omega \rangle$ and $\Pi (A)|\Omega \rangle =0$ iff $A=0.$ There exists the modular conjugation $J$ given by $J\Pi( A)|\Omega =\Pi( A^*)|\Omega \rangle .$ The modular conjugation defines a mapping $JAJ$ from the algebra into the commutant
$$JAJ \Pi (B)=\Pi(B)JAJ,\quad JABJ=JBJ JAJ.$$
In addition $JAJ|\Omega \rangle =\Pi A^*|\Omega \rangle .$

The tracial state is invariant under all approximately inner automorphisms so that every automorphism $\tau $ can be implemented by a unitary
$$ \Pi (\sigma A)=W\Pi (A) W^*, W|\Omega \rangle =|\Omega \rangle .$$
This unitary implements also the corresponding automorphism on the commutant  $J \tau AJ=\tau J A J.$ Altogether the matrix algebra at a point is mapped into a matrix algebra in the commutant and together with the shift  the commutant is also the representation of a quantum algebra on the lattice with the same dimension.
Consider $\tau _t$ to be a continuous automorphism group and assume that it is strong asymptotically abelian:
$$st\lim _{t\rightarrow \infty } \Pi( [\tau _t A,B])=0.$$
Then also $$st\lim _{t\rightarrow \infty }[J\tau _t AJ, JBJ]=0,\quad [J\tau _tAJ,\Pi (B)]=0  \quad [\Pi (A), (\tau _t),JBJ]=0$$ and therefore also
$$st\lim _{t\rightarrow \infty }[\tau _t(\Pi( A) JBJ, \Pi (C) JDJ]=0.$$
The automorphism group is strongly asymptotically on the whole algebra of the Hilbertspace. Assume $U$ is a unitary in $\A$. Then JUJ is again unitary. For strongly asymptotically abelian automorphism groups
$$
st\lim _t \Pi(\tau _t U)\Pi( A) \Pi(\tau _t(U^*)|\Omega \rangle =\Pi (A)|\Omega \rangle =st\lim _t\tau _t \Pi (U)JUJ \Pi (A)|\Omega \rangle .$$
Using that the representation is faithfull it follows that
$$st\lim _t\Pi (U) JUJ=1$$
and we have transferred the commutativity property of the commutant where two time depending operators are involved into the convergence property of  a single operator.
The algebra is built by combinations of the Weyl-operators $U_{r_1,r_2}$ at some point and their shifts. We restrict our observations to automorphism groups that commute with the shift. Therefore it suffices to examine
$st.\lim_t \tau _t (\Pi (U_{r_1,r_2}) JU_{r_1,r_2}J) $. Here $ \Pi(U_{r_1,r_2}) JU_{r_1,r_2}J$ can be considered as a Weyl-operator in the algebra $\M^d \otimes M^d$ of the form $U_{r_1,r_2}\otimes U_{-r_1, r_2}$ where we have taken into account that the modular conjugation implements an antiisomorphism. This unitary has eigenvalues $e^{i \frac{2\pi }{d}(n_1(r_1-r_2)+n_2(r_1 +r_2)}$.

 Every automorphism preserves the set of eigenvalues but can change their degeneracy, which for the operators we consider in our  representation is infinite. In the strong limit eigenvalues can disappear. Demanding that $st.\lim_{t\rightarrow \infty }\tau _t (\Pi (U_{r_1,r_2})JU_{r_1,r_2}J)=1$ is satified
if $st.\lim P_{r_1,r_2}(0)=1$ whereas $st.\lim_{t\rightarrow \infty } P_{r_1,r_2}(j)=0 \quad \forall j\neq 0$ where $P_{r_1,r_2}(j)$ is the projection of $\Pi(U_{r_1,r_2})JU_{r_1,r_2}J$ with eigenvalue $j.$ This has to hold for all $r_1,r_2$, therefore for the intersection of $P_{r_1,r_2}(0).$ With $$[U_{r_1,r_2}\otimes U_{-r_1,r_2}, U_{s_1,s_2}\otimes U_{-s_1,s_2 }]=0$$
all relevant projections commute  and the intersection of those with eigenvalue $O$ is given by $|\sum _j j\otimes j\rangle \langle \sum _j j\otimes j|$ with $|j\rangle $ an orthonormal basis in $M^d$ where the correspondence between the basis in the tensorproduct is fixed by the modular conjugation. The corresponding state is the maximally entangled state between the two matrices.

In the representation of the quasilocal algebra it is infinetely degenerate.  Especially for the shift $\sigma _y M_x=M_{x+y}$ we know that it is norm asymptotically abelian and therefore also strongly
 asymptotically abelian. Considering the projection as a projection in the total Hilbertspace located at the position $x=0$ it follows that $$st.\lim _{x\rightarrow \infty }\sigma _x |\sum _j j\otimes j\rangle _0\langle \sum _j j\otimes j|_0\otimes 1 =1$$
We collect the observations in
\begin{theorem} An automorphism group with $t\in Z$ or $t\in R$ is strongly asymptotically abelian if
$$st \lim _{t\rightarrow \infty}\tau _t |\sum _j j\otimes j \rangle _0 \langle \sum _j j\otimes j|_0 \otimes 1 =1$$\end{theorem}
Assume $t\in R$ and the automorphism is continuous with derivative $\delta$.  Since the convergence holds for every starting point it follows that also
$$\st.\lim \tau _{t\rightarrow \infty } \delta  |\sum _j j\otimes j\rangle _0 \langle \sum _j j\otimes j|_0 \otimes 1 =0,$$i.e. expressed by the relevant operators $$\lim _t e^{iHt}[H,P]e^{-iHt}=0.$$ Strong convergence holds also for the product, therefore in combination with the convergence of $P$ this implies
$$ \lim _t \tau _t(\delta P \delta P ) = \st.\lim _t \tau _t(P(H(1-P)H-H^2)P)=0.$$
This operators is not negative. It can converge strongly to $0$ only if it has eigenvalue $0$ and this eigenvalue is infinitely degenerate so that the corresponding projector can converge to $1.$ We have to control whether this can happen.

The Hamiltonian in its action as defining the derivative on the localized projection can be approximated by local operators $H_{\Lambda }\otimes 1-1\otimes H_{\Lambda }$. Defining the action of $H_{\Lambda }$  on the eigenvectors $|j\rangle _0$ in $M^d \otimes M^d $ as $H_j$
$$|\sum _j j\otimes j\rangle _0 \langle \sum _k k\otimes k| _0H_{\Lambda }(1-|\sum _l l\otimes l\rangle  _0\langle \sum _m m\otimes m|_0)H_{\Lambda }\sum _n n\otimes n\rangle _0 \langle \sum _p p\otimes p|_0=$$
$$|\sum _j j\otimes j\rangle \langle \sum _k (H_k k\otimes k -k\otimes H_k k|(1-|\sum _l l\otimes l\rangle \langle \sum _m m\otimes m|)\sum _n (H_n n\otimes n-n\otimes H_n n\rangle \langle \sum _p p\otimes p|=$$
$$|\sum _j j\otimes j\rangle \sum _k \sum _l\langle (H_k k\otimes k -k\otimes H_k k|(1-| k\otimes k\rangle \langle  l\otimes l|H_l l\otimes l-l\otimes H_l l\rangle \langle \sum _p p\otimes p|$$
The terms that contribute can be combined to

$$\sum _{k,l}\langle (H_k\otimes 1-1\otimes H_k)k\otimes k|(H_l\otimes 1-1\otimes H_l) l\otimes l\rangle$$
which reduces to
$$ \sum _{kl} \langle k|H_k^2|k\rangle - \langle k|H_k|l\rangle\langle l|H_k k\rangle =0$$
The operators are given by the local Hamiltonian reduced to their action on the eigenvectors$|k \rangle $, but they map into a larger Hilbertspace, since the Hamiltonian is not localized at the point $0$.
We can write $H_k$ in matrixnotation as $ H_k(k,l')$ where $k$ runs over $1,..d$ whereas $l'$ runs over a larger region. $H_k$ and $H_l$ are related by $ H_k(k,l)=H_l(k,l)$.
Therefore the above sum expressed in the matrix elements of $H_k$ reads
$$\sum _{k,l'} |H_k(k,l)|^2-\sum _{k,l}|H_k(k,l)=0=\sum _{k,l'>d}|H_{k,l'}|^2$$
and vanishes only if $H_k(k,l')=0 \forall l'>d.$ This reduces to Hamiltonians with interaction only inside the lattice point and do not give a time evolution that is asymptotically abelian. It follows:
\begin{theorem} Continuous automorphism groups that are the limit of local automorphism groups are never strongly asymptotically abelian and therefore not norm asymptotically abelian.\end{theorem}
We have formulated the observation in the representation given by the tracial state as far as this is sufficient to exclude norm asymptotical abelianess.  However we proved the stronger result, that it is not strongly asymptotically abelian. We only used that the representation is cyclic and faithful therefore admits a modular conjugation. Since this holds in every KMS-state with trivial center also in the GNS representations of KMS states continuous automorphism groups that are approximately inner cannot be strongly asymptotically abelian. This conclusion does not hold for the ground state or for other states with irreducible representation.
\section{Consequence for the shift}
For the quantum algebra on the lattice the shift is by construction norm asymptotically abelian. In \cite{RWW} the question was raised whether it allows  a continuous extension as it holds for the shift on the Fermi algebra. Based on two point interaction any construction as limit of local automorphisms fails. In \cite{MN} we started from the continuous extension possible for the Fermi algebra and showed that it cannot be transferred to the quantum algebra on the lattice.  The extension corresponded to a restriction that was not expressed as interaction but as a quasifree evolution, in some sense similar to the restriction in \cite{RWW}. In addition it was shown that
\begin{theorem} Assume that for a continuous  automorphism group $\alpha _t$ with $t\in R$
$$ norm-\lim _{n\rightarrow \infty}[\alpha _n (A),B]=0, n\in Z$$
then also
$$norm-\lim _{t\rightarrow \infty }[\alpha _t(A),B] =0 $$ \end{theorem}
The proof is based on splitting in small steps together with the continuity property.

Based on theorem 2 we can generalize the observations in \cite{RWW} and \cite{MN} to
\begin{theorem} Different to Fermi systems the shift on the quantum algebra on the lattice cannot be extended to a continuous automorphism group that can be locally approximated.\end{theorem}

\section{Localization and delocalization}
Classical dynamical systems are given by a probability space equipped with a finite probability measure $\mu$ and a continuous measurable map $T(t)$ for which the measure is invariant $\mu \circ T_t=\mu.$ The system is ergodic if the average over $f(T_tx)$ is $\mu$-almost constant. This can be improved by demanding that the average is replaced by the limit $t\rightarrow \pm \infty.$

For the quantum analog, the quantum algebra on the lattice, the finite probablity measure is replaced by a state invariant under the automorphism group $\tau _t$. The ergodic behaviour is determined by the spectral properties of the unitary $U _{\tau }(t)$ implementing the automorphism. For automorphisms that are norm asymptotically abelian this spectrum is apart from the eigenvalue $1$ on the eigenstate $|\Omega \rangle $ implementing the invariant state  absolutely continuous and according to Riemann-Lebesque guarantees that every vector $|\Psi \rangle $ converges strongly to
$\langle \Omega |\Psi \rangle |\Omega \rangle $ so that every locally perturbed state converges to the invariant state.

 For demanding that the invariant mean of the perturbed state coincides (up to normalization) with the invariant state strong convergence of the vectors is not necessary. Using the tracial property or the KMS property it suffices that $|\Omega \rangle $ is the only eigenstate to the eigenvalue $1$  so that operators converge in the mean weakly to c-numbers. Using the additivity of the spectrum that is a consequence of our assumption, that the continuous automorphism group commutes with space translation, it follows, that $1$ has to be the only eigenvalue and the rest of the spectrum has to be continuous. Expressed by the commutators this implies that at least in the mean $\eta -weak\lim _t \Pi([\tau _t A , B]=0 $. The delocalization does not happen in norm but it is sufficient to restrict correlations in the invariant state.

Evidently not all continuous automorphism groups satisfy this requirement. A counterexample is given by the Emch-Radin-model \cite {GGE}, \cite{Ra} where the Hamiltonian only includes $\sigma _z^j$ and therefore acts trivially on the subalgebra buit by $\sigma _z^j.$ However  continuous Fermi systems with Galilei invariant interaction \cite{NT2} are shown to be weakly asymptotically abelian. In \cite{NI} the proof was adapted to discrete Fermi systems with the necessary modification and the resulting restriction, that the pure point spectrum of $U_{\tau }$ could be shown to be empty but without specifying the remaining continuous spectrum.
We repeat the result on the pure point spectrum and examine it for examples of the quantum algebra on the lattice.

\begin{theorem} Let the continuous automorphism group be implemented by
$\sum _x \sigma _x H_{\Lambda }.$  Consider the automorphism $\gamma (g) \sigma _x A_0=e^{ig\sum _x n(x) \sigma _xB_0}\sigma _xA_0 e^{-ig\sum _x n(x)\sigma _xB_0}$ with $A_0$ localized at $x=0.$ Therefore the automorphism $\gamma (g)$ acts strictly local  but according to the choice of $n(x)$ with increasing strength with increasing $x$. Assume that $H_{\Lambda }$ and $B_0$ and $n(x)$ are related such that $\gamma (g) \tau _t \gamma (-g) =\tau _t(g)$ is again a continuous automorphis group in $t$ that commutes with the shift.   Then the time evolution in the tracial state has apart from the GNS vector $|\Omega \rangle $ a continuous spectrum. \end{theorem}
Proof: By assumption on the time evolution
\beq \sigma (x) \gamma (g) \tau (t) \gamma (-g)\sigma (-x)=\gamma (g) \tau (t) \gamma (-g) =\tau _g (t)\eeq
The tracial state is invariant under all automorphisms that can be locally approximated. Therefore these automorphisms are implemented by unitaries that let the GNS vector $|\Omega \rangle $ invariant. Assume $\tau (t)A|\Omega \rangle =e^{iEt}A|\Omega \rangle$ for some operator $A$ that can be approximated arbitrary well by a local operator. Choose $g$ such that $||\gamma (g)A -A|\Omega \rangle||< \epsilon .$ Then also $$||\tau (t) (\gamma(g)A-A)|\Omega \rangle ||=||\gamma (-g) \tau (t) (\gamma (g)A-A)|\Omega \rangle || =||\tau _g(t)A-e^{iEt}\gamma (-g)A|\Omega \rangle ||<2\epsilon.$$
Let $e^{iHt}$ implement the automorphism group $\tau _t $ and $e^{igG}$ implement the automorphism $\gamma (g).$ Then due to our assumption with $A|\Omega \rangle $ an eigenoperator to $e^{iEt}$ also $\sigma _xA|\Omega \rangle $ is an eigenvector to the same eigenvalue.
Therefore

$$||(e^{iHt}-e^{iEt})e^{ig\sigma _xG}A|\Omega \rangle ||\leq 2\epsilon \quad \forall x,t, g\leq g_0.$$

With increasing $x$ $\gamma (g)$ rotates locally with increasing velocity. Therefore varying over $t$ and $x$ it follows that $A$ does not only approximate an eigenvector for $H$ with eigenvalue $E$ but also an eigenvector for $G$ with eigenvalue $0.$ If $B_0$ has eigenvalues different from $0$ only $|\Omega \rangle $ satisfies this demand.

Expressed as a property of the time evolution on the quantum algebra it follows.
\begin{theorem} Under the assumption of Theorem (5) there are no local operators that remain strictly local in the course of time. \end{theorem}
Proof: For a unitary-group with only continuous spectrum every subspace that is invariant under the evolution is infinite dimensional. Local operators create a finite dimensional subspace. Therefore no vector respectively operator can stay in a finite dimensional subspace respectively local algebra.

It remains to look for examples that satisfy the requirements of the theorem. The examples for  the Fermi lattice of quasifree evolution combined with interaction involve only the even algebra. They can be mapped into evolutions on the quantum algebra on the lattice and serve as examples. Another example is given by

Example 1: The Heisenberg Fermi magnet in arbitrary dimensions with nearest neighbor interaction

The time evolution is implemented by $$\sum _{j,k; ||j-k||=1}(\sigma _x^j \sigma _x^k +\sigma _y^j \sigma _y^k +\sigma _z^j \sigma _z^k)$$
with $j,k$ indicating the position and $\sigma _{\alpha }$ the Pauli matrices.
For $B_0$ we choose $\sigma _z^0 $ and $n(j)=j .$ Then
$$\gamma (g)(\sigma _x^j \sigma _x^k)=(\cos(nj)\sigma _x^j + \sin (nj)\sigma _y^j)(\cos (nk)\sigma _x^k+ \sin (nk)\sigma _y^k).$$
Combining the different terms, taking into account that only nearest neighbors are involved with $||j-k||=1$ and using
$$\cos \alpha \cos \beta +\sin \alpha \sin \beta = \cos (\alpha -\beta )$$
the dependence of the rotation of the Pauli matrices on the location disappears and the assumption of the theorem is satisfied.

Example 2: Consider the contributions to the Hamiltonian
$$ |j\otimes k\rangle _x \langle k \otimes j|_y + |k\otimes j\rangle _x\langle j\otimes k |_y$$
again restricting to nearest neighbors and $ |j\rangle $ and $|k\rangle $ orthogonal vectors for $M^d$. Taking $$B_0=|j\rangle \langle j|-|k\rangle \langle k| $$ the same calculation confirms that $\gamma _g\tau _t \gamma (-g)$ is space translation invariant.

Otherwise for the Hamiltonian given by $|j\otimes k\rangle \langle j\otimes k|+h.c.$ the dependence on the location does not cancel.  One has to look for another idea to prove or disprove the absence of the pure point spectrum.

Even if we can control that the pure point spectrum apart from the GNS-vector is empty there  remains as open problem whether the continuous spectrum is absolutely continuous as we are used to in few particle quantum theory. If so we could apply
Riemann-Lebesgue and could conclude that in the tracial state the continuous automorphism group is weakly asymptotically abelian.  A further argument on the control of the continuous spectrum would be necessary. Notice also that we assumed that $\gamma (g)$ is implemented by a unitary. This is only given if we are in the tracial state. Generalizations to KMS states would ask for additional assumptions and additional estimates. However this does not effect Theorem (6). From this viewpoint the result of \cite{LR} is improved. But an explicit lower bound for the possible velocity estimated in \cite{LR} is not available so far.

\section{Conclusion}
We have studied properties of continuous automorphism groups on the quantum algebra on the lattice. We have shown, that norm asymptotical abelianess on the total algebra is not possible. However we have given no statement about subalgebras. With the XY-model \cite{A} an example exists, where the continuous evolution is norm asymptotically abelian for the even subalgebra. But our proof indicates, that the assumption for such a subalgebra respectively for such an automorphism group is rather restrictive.

All considerations were done in the GNS-representation given by the tracial state. In this representation the quasilocal structure is inherited by all bounded operators in the Hilbert space and offers the possibility to control convergence on appropriately chosen projectors that do not belong to the observable algebra. The spectrum of the unitary implementing the automorphism group was shown to be continuous with the exception of the GNS-vector which is an eigenvector to eigenvalue $1$ for all unitaries implementing an automorphism.

As an interesting byproduct it followed that the shift cannot be extended to a continuous automorphism group, differently as for Fermi systems.

The absence of norm asymptotic abelianess  implies that local observables do not totally escape to infinity but keep some local algebraic relations. Nevertheless they do not stay localized therefore they also have algebraic relations with infinity. How far this spreading has consequences for the ergodic properties of the system and give indications for the may be limited existence of invariant states apart from KMS states remains as open problem.

\bibliographystyle{plain}

\end{document}